\documentclass[aps,pra,twocolumn,superscriptaddress,showpacs]{revtex4}
\usepackage{amsfonts}
\usepackage{graphics}
\usepackage{amsmath}
\usepackage{amssymb}
\usepackage{graphicx}
\usepackage{color}

\begin{document}

\title{Minimal qudit code for a qubit in the phase damping channel}
\author{Stefano Pirandola}
\affiliation{Research Laboratory of Electronics, Massachusetts Institute of Technology,
Cambridge, MA 02139, USA}
\author{Stefano Mancini}
\affiliation{Dipartimento di Fisica, Universit\`{a} di Camerino, I-62032 Camerino, Italy.}
\author{Samuel L. Braunstein}
\affiliation{Department of Computer Science, University of York, York YO10 5DD, United
Kingdom}
\author{David Vitali}
\affiliation{Dipartimento di Fisica, Universit\`{a} di Camerino, I-62032 Camerino, Italy.}
\date{\today}

\begin{abstract}
Using the stabilizer formalism we construct the minimal code into a $D$-dimensional Hilbert space (qudit) to protect a qubit against phase damping.
The effectiveness of this code is then studied by means of input-output
fidelity.
\end{abstract}

\maketitle

\section{Introduction}

\label{sec:int}

Quantum Error Correction (QEC) theory \cite{S95,CS96,G96,KLV00} concerns the
possibility to protect against environmental noise when storing or
transmitting quantum information. This possibility relies, likewise in the
classical error correction theory, in \emph{redundancy}. This implies
embedding a quantum information unit (a qubit) belonging to a given Hilbert
space ($\mathcal{H}_{2}$) into a larger one ($\mathcal{H}_{D}$ with $D>2$).
The latter is usually chosen as $n$ times the tensor product of the former,
so that $D=2^{n}$ and $\mathcal{H}_{D}=\mathcal{H}_{2}^{\otimes n}$. We
generally refer to this kind of encoding%
\begin{equation}
\mathcal{C}:\mathcal{H}_{2}\longrightarrow \mathcal{H}_{2}^{\otimes n}~,
\label{Block}
\end{equation}%
as \emph{block-encoding} of the qubit. However, it is obvious that
whatever $\mathcal{H}_{D}$ (with $D>2$), extra state space is available
and could potentially be exploited without any restriction. As a
consequence, the alternative possibility is to embed a qubit into a $D$%
-dimensional quantum system, i.e., a qudit, where $D$ can be made even
infinite (in the limit where the qudit becomes a quantum oscillator, i.e., a
bosonic mode \cite{CVSamSeth,GKP01}). We refer to this kind of encoding%
\begin{equation}
\mathcal{C}:\mathcal{H}_{2}\longrightarrow \mathcal{H}_{D}\neq \mathcal{H}%
^{\otimes n}~,  \label{Graft}
\end{equation}%
as to \emph{qudit-encoding} of the qubit. In a standard QEC framework, such
encodings are coupled to suitable decoding stages, where a recovery
operation (e.g., syndrome extraction and error correction) restores the
original quantum information by removing the (correctable) errors induced by
the noisy action of the environment.

Note that the encodings of Eqs.~(\ref{Block}) and~(\ref{Graft}) are not
equivalent. A first simple reason relies in the available dimensions for
qudit-encoding, which are not necessarily restricted to powers of $2$.
Moreover, at a more fundamental level, the errors affecting the two storing
systems are different, that is, they form two different algebras. For a
block of qubits errors are given by combinations of bit and phase flips,
which are representable in terms of products of Pauli matrices. A single
qudit instead is affected by amplitude and phase shifts (implying that,
asymptotically, a single bosonic mode is affected by diffusion in position
and momentum \cite{GKP01}), which are represented by the unitarily
generalized Pauli matrices for a single qudit.

Pioneering advances in the QEC\ with higher-dimensional spin systems \cite%
{Knill,QuditCodes} and bosonic modes \cite{CVSamSeth} were achieved during
the nineties. More recently, Ref.~\cite{GKP01} introduced novel kinds of
codes for qudits, known as \emph{shift-resistant} (SR) quantum codes. In its
simplest formulation, a SR\ code corresponds to embedding a logical qubit
into a larger qudit, followed by a recovery stage which restores the quantum
information from a bounded set of quantum errors (i.e., amplitude and phase
shifts whose \emph{weight} is less than some critical value). In particular,
Ref.~\cite{GKP01} showed that a qudit of dimension $D=18$ represents the
smallest quantum system able to protect a logical qubit from a single
quantum error, where the corresponding $5$-qubit block code $[[5,1,3]]$ of
Ref.~\cite{Fivequbits} needs a Hilbert space of dimension $D=2^{5}>18$. Let
us underline that both of these codes are stabilizer codes \cite{G96} and are \emph{%
perfect}, roughly meaning that they need minimal quantum resources for their
task \cite{note1}.

The latter peculiarity is very important since the primary issue for having
experimentally feasible QEC codes consists in simplifying their complexity.
In fact, the importance of using minimal resource codes relies on our
current difficulty in performing high fidelity operations on a small number
of qubits \cite{Exp}. However, apart from the optimality of the above
perfect codes (which are designed to defeat general quantum errors), it is
still an open problem to find the most efficient quantum codes which enable
QEC within \emph{specific error models}. In fact, if the dominant
decoherence process in a physical system is of a specific nature and well
known, one can look for a corresponding quantum error correction scheme
whose quantum complexity is as small as possible. Such a problem has been
raised, for the first time, in Ref.~\cite{SamPhase} for protecting logical
qubits against dephasing. Later, Ref.~\cite{Chuang} proposed an optimal code
embedding a qubit in a block of bosonic modes able to protect against the
effect of amplitude damping.

To date, nobody has analyzed the same problem for qudit-encoding, i.e.,
nobody has considered the engineering of a minimal single-qudit code able to
protect a logical qubit against a specific kind of decoherence.{\ Only Ref.~%
\cite{AMM06} pointed out that qudit-encoding is not effective by itself when specific error models are taken into account. That is, without a
suitable error correction (recovery) operation, the extra space cannot be exploited to protect against errors.} In this paper, we consider the
qudit-encoding in a QEC\ framework (i.e., with a suitable recovery stage) and we design the minimal\ codes which are able to protect a logical
qubit against a single class of errors, such as amplitude or phase shifts. Note that we are here considering minimal codes which are
\emph{quasi-classical}. In fact, even if they encode quantum information (one logical qubit), the environmental error-model here is
\emph{classical}, in the sense that the correctable errors occur only in a preferred basis. The two complementary bases of a single qudit are
perfectly symmetric, being connected by a discrete Fourier transformation. Therefore, by fixing the unperturbed basis (pointer basis) to be the
computational one, we can always define as phase-damping the damping that affects the complementary basis. We will show the robustness of a
minimal qudit code in preserving the encoded quantum information against this kind of error.

The layout of the paper is the following. In Section \ref{sec:cod} we
present the code's construction and its performance against shift-errors.
Section \ref{sec:pha} is devoted to the phase damping channel. Section \ref%
{sec:fid} shows the performance of the code against phase damping in terms
of input-output fidelity. Finally, Section \ref{sec:con} is the conclusion.

\section{The code}

\label{sec:cod}

\subsection{Qudits}

Let us consider a qudit, i.e., a $D$-dimensional spin-system. In its Hilbert
space $\mathcal{H}_{D}$ we choose a computational basis $\{|j\rangle \}$
labeled by modular integers $j\in \mathbb{Z}_{D}:=\{0,\ldots ,D-1\}$. An
arbitrary unitary transformation $\mathcal{H}_{D}\rightarrow \mathcal{H}_{D}$
can be expanded in terms of $D^{2}$ \emph{generalized} Pauli operators \cite%
{note2}%
\begin{equation}
X^{a}Z^{b}~,\quad a,b\in \mathbb{Z}_{D}~,  \label{QEC_PAULI_gen}
\end{equation}%
which are defined by
\begin{equation}
X\left\vert j\right\rangle =\left\vert j\oplus 1\right\rangle ~,~Z\left\vert
j\right\rangle =\omega ^{j}\left\vert j\right\rangle ~,  \label{Pauli_DEF}
\end{equation}%
where $j_{1}\oplus j_{2}:=j_{1}+j_{2}(\mathrm{mod}D)$ and%
\begin{equation}
\omega :=\exp (i2\pi /D)~.
\end{equation}%
Such unitary operators satisfy the anticommutation relation
\begin{equation}
ZX=\omega XZ~,  \label{Anticomm}
\end{equation}%
and their eigenstates are connected by%
\begin{equation}
\widetilde{\left\vert i\right\rangle }=\sum_{j=0}^{D-1}H_{ij}\left\vert
j\right\rangle ~,  \label{altrabase}
\end{equation}%
where $X\widetilde{\left\vert i\right\rangle }=\omega ^{j}\widetilde{%
\left\vert i\right\rangle }$, and $H$ is the $D\times D$ Fourier matrix with
entries
\begin{equation}
H_{ij}:=\frac{\omega ^{-ij}}{\sqrt{D}}~,\quad i,j\in \mathbb{Z}_{D}~.
\label{Had_Generic}
\end{equation}%
Accordingly, a general quantum error acting on the qudit can be decomposed
in the error basis of Eq.~(\ref{QEC_PAULI_gen}). Its elements, i.e., the
generalized Pauli operators, represent the basic quantum errors which a
quantum correcting code must correct. According to Eq.~(\ref{Pauli_DEF})
these are distinguished as amplitude shifts $X^{a}$ and complementary phase
shifts $Z^{b}$. Multiplying by suitable phase factors $\omega ^{j}$ the
elements of Eq.~(\ref{QEC_PAULI_gen}), one defines the qudit Pauli group
and, consequently, extends the stabilizer formalism \cite{G96,GKP01}. These two
kinds of errors, if considered separately, represent an abelian group and
their correction can thus be performed through quasi-classical codes.

\subsection{Single errors}

It is natural to ask what is the smallest $D$-level system which protects an
encoded qubit from a single amplitude shift $X^{\pm 1}$. Let us first
consider an example with $D=6$, so that $\omega =\exp \left( i\pi /3\right) $%
. A logical qubit can be encoded in the two codewords stabilized by the
generator $Z^{2}$, i.e.,
\begin{equation}
\overline{\left\vert 0\right\rangle }:=\left\vert 0\right\rangle ~,~%
\overline{\left\vert 1\right\rangle }:=\left\vert 3\right\rangle ~,
\end{equation}%
where
\begin{equation}
Z^{2}\left\vert 0\right\rangle =\left\vert 0\right\rangle ~,~Z^{2}\left\vert
3\right\rangle =\left\vert 3\right\rangle ~.
\end{equation}%
In such a case the measurement of the stabilizer preserves every coherent
superposition $\left\vert \varphi \right\rangle =\alpha \left\vert
0\right\rangle +\beta \left\vert 3\right\rangle $, i.e.,%
\begin{equation}
Z^{2}\left\vert \varphi \right\rangle =\left\vert \varphi \right\rangle ~,
\end{equation}%
while it detects single $X$ errors, i.e.,%
\begin{equation}
Z^{2}X^{\pm 1}\left\vert \varphi \right\rangle =\omega ^{\pm 2}X^{\pm
1}\left\vert \varphi \right\rangle ~.
\end{equation}%
Alternately, one must consider the complementary generator $X^{2}$ for
correcting single{\ $Z^{\pm 1}$} errors, i.e., one must encode the qubit
into the codewords%
\begin{equation}
\overline{\left\vert +\right\rangle }:=\widetilde{\left\vert 0\right\rangle }%
~,~\overline{\left\vert -\right\rangle }:=\widetilde{\left\vert
3\right\rangle }~,
\end{equation}%
where
\begin{equation}
X^{2}\widetilde{\left\vert 0\right\rangle }=\widetilde{\left\vert
0\right\rangle }~,~X^{2}\widetilde{\left\vert 3\right\rangle }=\widetilde{%
\left\vert 3\right\rangle }~.
\end{equation}%
According to Eq.~(\ref{altrabase}), one can express these codewords in the
computational basis as%
\begin{equation}
\widetilde{\left\vert 0\right\rangle }=\frac{1}{\sqrt{6}}\sum%
\limits_{j=0}^{5}\left\vert j\right\rangle ~,~\widetilde{\left\vert
3\right\rangle }=\frac{1}{\sqrt{6}}\sum\limits_{j=0}^{5}(-1)^{j}\left\vert
j\right\rangle ~.
\end{equation}

\subsection{Multiple errors}

\label{sec:multiple}

\begin{figure}[tbph]
\vspace{-0.5cm}
\par
\begin{center}
\includegraphics[width=0.5\textwidth]{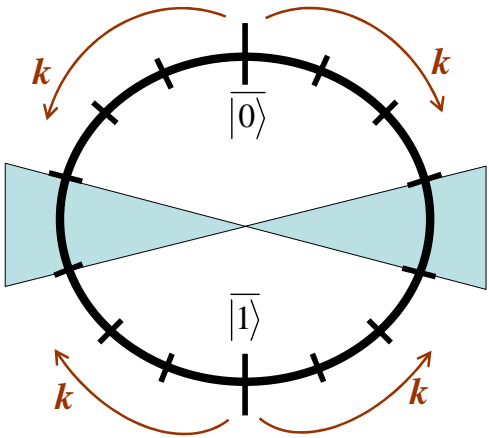}
\end{center}
\par
\vspace{-0.5cm}
\caption{Pictorial view of errors' effect on code states. The shadowed
area separates the correctable error spaces associated with the two codewords.}
\label{clock}
\end{figure}

From the previous example we argue that in order to correct $k$ shifts we
need a qudit with
\begin{equation}
D=4k+2  \label{dimension}
\end{equation}%
levels. This can be understood by means of the \textquotedblleft
clock\textquotedblright\ picture of Fig.~\ref{clock}, and is easily proven in
the following.
Consider, for example, the case of $k$\ amplitude shifts $X^{\pm k}$, but
the reasoning is perfectly symmetric for the complementary errors. In order
to determine two possible codewords, we must consider the eigenvalue equation%
\begin{equation}
Z^{G}\left\vert j\right\rangle =\omega ^{jG}\left\vert j\right\rangle ~,
\label{omegaG}
\end{equation}%
where $j\in \mathbb{Z}_{D},$ $0\neq G\in \mathbb{Z}_{D}$ and $D$ must be
determined. In Eq.~(\ref{omegaG}), the state $\left\vert j\right\rangle $ is
stabilized if and only if $\omega ^{jG}=1$, and this happens in either the
trivial case $j=0$ or in the case
\begin{equation}
jG=D~.  \label{jG}
\end{equation}%
Note that, since $j<D$, we must necessarily set $G\geq 2$ for the weight of
the $Z$-generator. The resulting code will be able to correct{\ all errors $%
X^{0},X^{\pm 1},X^{\pm 2},\ldots X^{\pm k}$} if and only if the stabilizer
generator $Z^{G}$ will commute with $X^{d}$, where $d:=2k+1$ defines the
distance of the code. Since%
\begin{equation}
Z^{G}X^{d}=\omega ^{dG}X^{d}Z^{G}~,
\end{equation}%
this will happen if and only if%
\begin{equation}
dG=D~,  \label{dG}
\end{equation}%
i.e., if the weight of the generator corresponds to the the ratio between
the dimension of the qudit and the code's distance. By comparing Eqs.~(\ref%
{jG}) and~(\ref{dG}) we must conclude that
\begin{equation}
j=d=2k+1~.  \label{j_value}
\end{equation}%
Then, by making the \textquotedblleft minimal\textquotedblright\ choice $G=2$
in Eq.~(\ref{jG}), we get the minimal dimension $D$ of Eq.~(\ref{dimension}%
). In conclusion, $k$ amplitude-shifts are corrected by enconding a qubit
into the codewords%
\begin{equation}
\overline{\left\vert 0\right\rangle }:=\left\vert 0\right\rangle ~,~%
\overline{\left\vert 1\right\rangle }:=\left\vert 2k+1\right\rangle ~,
\label{codewords_Z}
\end{equation}%
of a $(4k+2)$-dimensional qudit. These codewords are stabilized by the
operator $Z^{2}$ and are connected by the logical flip gate $\bar{X}%
:=X^{2k+1}$. Analogously, $k$ phase-shifts are corrected by means of the
codewords%
\begin{eqnarray}
\overline{\left\vert +\right\rangle } &:&=\widetilde{\left\vert
0\right\rangle }=\frac{1}{\sqrt{4k+2}}\sum\limits_{j=0}^{4k+1}\left\vert
j\right\rangle ~,  \label{zero_Z} \\
\overline{\left\vert -\right\rangle } &:&=\widetilde{\left\vert
2k+1\right\rangle }=\frac{1}{\sqrt{4k+2}}\sum\limits_{j=0}^{4k+1}(-1)^{j}%
\left\vert j\right\rangle ~,  \label{uno_Z}
\end{eqnarray}%
which are stabilized by $X^{2}$ and are connected by the logical phase gate $%
\bar{Z}:=Z^{2k+1}$. Both these codes are perfect, since the correctable
error spaces (of dimension $d$) associated to their codewords just barely
fit in the qudit space (of dimension $D=2d$). We may refer to these codes as
to \emph{minimal amplitude code} $\{\left\vert 0\right\rangle ,\left\vert
2k+1\right\rangle \}$ and \emph{minimal phase code} $\{\widetilde{\left\vert
0\right\rangle },\widetilde{\left\vert 2k+1\right\rangle }\}$, respectively.
It is clear that they are equivalent up to a (discrete) Fourier
transformation.

Note that, correspondingly, the minimal qubit block-code which is able to correct $k$ phase (or amplitude) error flips works via majority voting
and, therefore, needs a block of $2k+1$ qubits. This is equivalent to considering a Hilbert space of dimension $D=2^{2k+1}$ which is exponential
rather than polynomial in $k$. This means that the qubit code is exponentially more demanding than the corresponding shift-resistant qudit code
at given weight $k$.

\subsection{Syndrome extraction and error recovery}

\subsubsection{Amplitude errors}

Consider an arbitrary coherent superposition of orthogonal codewords of
the amplitude code $\{\left\vert 0\right\rangle ,\left\vert
2k+1\right\rangle \}$, i.e.,%
\begin{equation}
\left\vert \varphi (0)\right\rangle =\alpha \left\vert 0\right\rangle +\beta
\left\vert 2k+1\right\rangle ~.  \label{encoded_state}
\end{equation}%
Suppose that an amplitude shift-error $X^{s}$, with \emph{syndrome} $-k\leq
s\leq k$, occurs on this superposition. Then, the logical state of Eq.~(\ref%
{encoded_state}) becomes%
\begin{equation}
\left\vert \varphi (s)\right\rangle :=X^{s}\left\vert \varphi
(0)\right\rangle =\alpha \left\vert 0\oplus s\right\rangle +\beta \left\vert
2k+1\oplus s\right\rangle ~.  \label{corruptedstate}
\end{equation}%
According to Sec.~\ref{sec:multiple}, such an error is detected by measuring
the complementary generator $Z^{2}$. In fact, this measurement gives%
\begin{equation}
Z^{2}\left\vert \varphi (s)\right\rangle =\omega ^{2s}\left\vert \varphi
(s)\right\rangle ~,
\end{equation}%
i.e., the syndrome is unambiguously extracted via the eigenvalue $\omega
^{2s}$ of $Z^{2}$ (non-degeneracy of the code), while the corrupted state is
preserved in the process. In order to realize this kind of quantum
non-demolition measurement, we must append an ancillary system to the
signal, let the joint system evolve according to a suitable unitary
interaction, and finally measure the ancilla. Since we must distinguish $%
2k+1 $ orthogonal errors ($k$ positive shifts, $k$ negative shifts, and the
no-shift), we need an ancillary system having at least $2k+1$ orthogonal
states, that we label by $\left\vert l\right\rangle _{A}$ with $l\in\mathbb{Z}$, $\left\vert
l\right\vert \leq k$.

In detail this correction process goes as follows. Let us introduce the $%
2k+1$ projectors
\begin{equation}
P(s):=\left\vert 0\oplus s\right\rangle \left\langle 0\oplus s\right\vert
+\left\vert 2k+1\oplus s\right\rangle \left\langle 2k+1\oplus s\right\vert ~,
\label{proj}
\end{equation}%
and construct the following unitary operation (generalized CNOT)
\begin{equation}
N:=\sum\limits_{s=-k}^{k}P(s)X_{A}^{s}~.  \label{Control_Operation}
\end{equation}%
It is then easy to check that $N$ realizes the syndrome extraction. In fact,
its effect on the joint system signal-plus-ancilla corresponds to leave the
corrupted state unchanged while shifting the ancilla by a quantity equal to
the syndrome, i.e.,%
\begin{equation}
N\left( \left\vert \varphi (s^{\prime })\right\rangle \otimes \left\vert
0\right\rangle _{A}\right) =\left\vert \varphi (s^{\prime })\right\rangle
\otimes \left\vert s^{\prime }\right\rangle _{A}~.  \label{final_state}
\end{equation}%
At this point, the measurement of the ancilla provides the syndrome $%
s^{\prime }$ and one restores the original signal state by applying the
corresponding inverse operator $X^{-s^{\prime }}$ to $\left\vert \varphi
(s^{\prime })\right\rangle $.

It is known that the last procedure, i.e., the error correction stage, can
be also implemented in a unitary manner. In fact, we can define
the correction operator%
\begin{equation}
C:=\sum\limits_{s=-k}^{k}X^{-s}\left\vert s\right\rangle _{A}\left\langle
s\right\vert ~,  \label{Correction}
\end{equation}%
and applying it to to the final state of Eq.~(\ref{final_state}). In such a
way we get%
\begin{equation*}
C\left( \left\vert \varphi (s^{\prime })\right\rangle \otimes \left\vert
s^{\prime }\right\rangle _{A}\right) =\left\vert \varphi (0)\right\rangle
\otimes \left\vert s^{\prime }\right\rangle _{A}~,
\end{equation*}%
thus recovering the initial encoded state of Eq.~(\ref{encoded_state}). Note
that the two unitary operators of Eqs.~(\ref{Control_Operation}) and~(\ref%
{Correction}) can be compacted together in a unique recovery operator
\begin{equation}
R:=CN=\sum\limits_{r,s=-k}^{k}X^{-r}P(s)\otimes \left\vert r\right\rangle
_{A}\left\langle r\right\vert X_{A}^{s}~.
\end{equation}%
The above derivation simply shows how the recovery procedure works properly
when a logical state $\left\vert \varphi (0)\right\rangle $ is affected by
amplitude error-shifts which are correctable, i.e., which fall within the distance
of the code. More generally, we may ask how the recovery works when such
errors are not necessarily correctable. This is a question that must be
answered if we want to test these codes in a quantum communication scenario
where the decoherence of a channel can be very strong. To this purpose we
must first derive the effect of recovery on a completely arbitrary state of
the qudit.

Thus, let us consider an arbitrary state
\begin{equation}
\rho =\sum_{i,j=0}^{D-1}\rho _{ij}\left\vert i\right\rangle \left\langle
j\right\vert ,\quad \rho _{ij}:=\langle i|\rho |j\rangle ~,  \label{rhoS}
\end{equation}%
of a qudit with dimension $D=4k+2$. The joint action of the recovery
operator reads%
\begin{gather}
R\left( \rho \otimes \left\vert 0\right\rangle _{A}\left\langle 0\right\vert
\right) R^{\dagger } \notag \\
=\sum\limits_{s,s^{\prime }=-k}^{k}X^{-s}P(s)\;\rho \;P^{\dagger }(s^{\prime
})X^{s^{\prime }}\otimes \left\vert s\right\rangle _{A}\left\langle
s^{\prime }\right\vert ~.
\end{gather}%
If we now trace out the ancilla, we get the \emph{recovery map} acting on the
qudit state
\begin{equation}
\mathcal{E}_{R}(\rho )=\sum\limits_{s=-k}^{k}X^{-s}P(s)\;\rho \;P^{\dagger
}(s)X^{s}~.
\end{equation}%
By virtue of Eqs.~(\ref{proj}) and (\ref{rhoS}), it is equal to%
\begin{gather}
\mathcal{E}_{R}(\rho )=\Phi (0,0)\left\vert 0\right\rangle \left\langle
0\right\vert +\Phi (0,2k+1)\left\vert 0\right\rangle \left\langle
2k+1\right\vert  \notag \\
+\Phi (2k+1,0)\left\vert 2k+1\right\rangle \left\langle 0\right\vert  \notag
\\
+\Phi (2k+1,2k+1)\left\vert 2k+1\right\rangle \left\langle 2k+1\right\vert ~,
\label{sums}
\end{gather}%
where
\begin{equation}
\Phi (x,y):=\sum\limits_{s=-k}^{k}\rho _{x\oplus s,y\oplus s}~.  \label{FI}
\end{equation}

\subsubsection{Phase errors}

For correcting phase errors the procedure is perfectly analogous to the
previous one. It is sufficient to exchange the role of $X$ and $Z$ and
account for the rotated codewords of the phase code $\{\widetilde{\left\vert
0\right\rangle },\widetilde{\left\vert 2k+1\right\rangle }\}$. {Thus, the
action of recovery on an arbitrary state $\rho $\ of the system is now
described by the map}
\begin{gather}
\mathcal{\tilde{E}}_{R}(\rho )=\tilde{\Phi}(0,0)\widetilde{\left\vert
0\right\rangle }\widetilde{\left\langle 0\right\vert }+\tilde{\Phi}(0,2k+1)%
\widetilde{\left\vert 0\right\rangle }\widetilde{\left\langle
2k+1\right\vert }  \notag \\
+\tilde{\Phi}(2k+1,0)\widetilde{\left\vert 2k+1\right\rangle }\widetilde{%
\left\langle 0\right\vert }  \notag \\
+\tilde{\Phi}(2k+1,2k+1)\widetilde{\left\vert 2k+1\right\rangle }\widetilde{%
\left\langle 2k+1\right\vert }~,  \label{Rec_Channel_PhaseOLD}
\end{gather}%
with
\begin{equation}
\tilde{\Phi}(x,y):=\sum\limits_{s=-k}^{k}\tilde{\rho}_{x\oplus s,y\oplus s}~,
\label{FI_tilde}
\end{equation}%
and $\tilde{\rho}_{ij}:=\widetilde{\left\langle i\right\vert }\rho
\widetilde{\left\vert j\right\rangle }$. In order to express these formulae
in the computational ($Z$) basis, we apply Eq.~(\ref{altrabase}) yielding
\begin{equation}
\tilde{\rho}_{ij}=\sum\limits_{l,m=0}^{D-1}H_{il}^{\ast }\rho _{lm}H_{mj}=%
\frac{1}{D}\sum\limits_{l,m=0}^{D-1}\rho _{lm}\omega ^{il-mj}~.
\label{Had_coordinates}
\end{equation}%
Therefore,%
\begin{equation}
\tilde{\Phi}(x,y)=\frac{1}{D}\sum\limits_{l,m=0}^{D-1}\rho
_{lm}(-1)^{xl-ym}\Delta (l-m,D)~,  \label{FI_Tilde_CompBase}
\end{equation}%
where%
\begin{equation}
\Delta (l-m,D):=\sum\limits_{s=-k}^{k}\omega ^{s(l-m)}=\frac{\sin\frac{\pi (l-m)}{2}}{\sin\frac{\pi (l-m)}{D}} \label{Kernel}
\end{equation}%
and $D=4k+2$ as usual. The formula of Eq.~(\ref{Rec_Channel_PhaseOLD}) is
crucial for our purposes. In fact, it will enable us to test the correcting
performance of our minimal phase-code $\{\widetilde{\left\vert
0\right\rangle },\widetilde{\left\vert 2k+1\right\rangle }\}$ in a quantum
communication scenario where the prevalent effect of decoherence is
ascribable to phase damping.

\section{Phase damping channel for qudits}

\label{sec:pha}

The phase damping (or phase flip) channel for a qubit can be defined by the following Kraus decomposition \cite{nielsen}
\begin{equation}
\mathcal{E}(\rho )=\sum_{i=0}^{1 }E_{i}\rho E_{i}^{\dagger }~, \label{Kraus_map2} \end{equation} with Kraus operators
\begin{equation}
E_{0}=\sqrt{\frac{1+\eta}{2}}I ~,~E_{1}=\sqrt{\frac{1-\eta}{2}}Z, \label{krausNC}
\end{equation}
where $I$ is the two-dimensional identity operator and $Z$ is given by Eq.~(\ref{Pauli_DEF}) with $D=2$. One can describe the phase damping
channel in an equivalent way, by adopting two different Kraus operators, related to those of Eq.~(\ref{krausNC}) by a unitary transformation
\begin{equation}
\mathcal{E}(\rho )=E_{0}^{\prime }\rho E_{0}^{\prime \dagger }+E_{1}^{\prime }\rho E_{1}^{\prime \dagger }~,  \label{PDamp_forQubits}
\end{equation}%
where now
\begin{equation}
E_{0}^{\prime }=\left\vert 0\right\rangle \left\langle 0\right\vert +\eta
\left\vert 1\right\rangle \left\langle 1\right\vert ~,~E_{1}^{\prime }=\sqrt{%
1-\eta ^{2}}\left\vert 1\right\rangle \left\langle 1\right\vert ~.
\end{equation}
The two Kraus decompositions of the phase damping channel for qubits, Eq.~(\ref{Kraus_map2}) and Eq.~(\ref{PDamp_forQubits}), suggest two
different generalizations to the general case of dimension $D$. The decomposition of Eq.~(\ref{Kraus_map2}) can be straightforwardly generalized
as
\begin{equation}
\mathcal{E}(\rho )=\sum_{m=0}^{D-1 }E_{m}\rho E_{m}^{\dagger }~, \label{Kraus_mapD} \end{equation} with
\begin{equation}
E_{m}=\sqrt{\left(\begin{array}{c}
D-1\\
m
\end{array}\right)
\left(\frac{1-\eta}{2}\right)^m \left(\frac{1+\eta}{2}\right)^{D-1-m}} Z^m ,\label{krausDweyl}
\end{equation}
which can be seen as a particular example of a Weyl channel \cite{Fuk05}, which is generally defined as
\begin{equation}
\rho\mapsto\mathcal{E}(\rho)=\sum_{m,n=0}^{D-1}\pi_{m,n}\, \left(Z^nX^m\right)\rho \left(X^m Z^n\right)^{\dag},
\end{equation}
with $0\le\pi_{m,n}\le1$ such that $\sum_{m,n=0}^{d-1}\pi_{m,n}=1$.

There is a however a different way to define the phase damping channel for a $D$-dimensional spin system, which is more closely related to the
usual physical meaning of phase damping. In fact, phase damping usually means that decoherence affects the elements of a given basis leaving the
elements of the complementary basis unchanged. As a consequence, in the basis of the unaffected states, decoherence destroys only the
off-diagonal elements of the density matrix of the qudit state. (It is evident that here we are conventionally fixing the elements $\left\vert
j\right\rangle $ of the computational $Z$-basis as the unchanged ones, but it is understood that a complementary damping channel can be
symmetrically defined).

The second definition of the phase damping channel for a qudit is suggested by Ref.~\cite{LOMI04} that has shown that the Kraus decomposition of
Eq.~(\ref{PDamp_forQubits}) is equivalent to another Kraus decomposition,
\begin{equation}
\mathcal{E}(\rho )=\sum_{i=0}^{\infty }E_{i}\rho E_{i}^{\dagger }~, \label{Kraus_mapinf}
\end{equation}
with an \emph{infinite} number of $E_i$, which in the case of qubits, are given by
\begin{equation}
E_{i}=\delta _{i0}\left\vert 0\right\rangle \left\langle 0\right\vert +\frac{%
\eta (-2\ln \eta )^{i/2}}{\sqrt{i!}}\left\vert 1\right\rangle \left\langle 1\right\vert ~.
\end{equation}%
The straightforward generalization of such channel in $D$ dimensions has been already studied in Ref.~\cite{AMM06} and has Kraus
operators
\begin{equation}
E_{i}:=\sum_{j=0}^{D-1}\frac{\left[ j\sqrt{-2\ln \eta }\right] ^{i}\eta ^{j^{2}}}{\sqrt{i!}}\left\vert j\right\rangle \left\langle j\right\vert
~. \label{E_i}
\end{equation}%
In all these examples, the parameter $\eta \in \lbrack 0,1]$ describes the strength of the damping \cite{AMM06}. Such a parameter can be assumed
independent from $D$ because it usually depends only on the bath characteristics, e.g., in the Markov approximation it depends upon the spectral
density of the bath only \cite{qnoise}. One can parameterize $\eta =e^{-\gamma }$, where $\gamma $ is proportional to the probability of a
phase error, so that phase damping is larger for smaller $\eta $. In particular, $\eta \rightarrow 1^{-}$ and $\eta \rightarrow 0^{+}$
correspond to the weak and strong damping limit, respectively.

It is easy to check from Eqs.~(\ref{Kraus_mapinf}) and~(\ref{E_i}), that an arbitrary density operator $\rho =\sum_{i,j=0}^{D-1}\rho
_{ij}\left\vert i\right\rangle \left\langle j\right\vert $ is mapped into the output density
operator given by%
\begin{equation}
\mathcal{E}(\rho )=\sum_{i,j=0}^{D-1}\rho _{ij}\eta ^{(i-j)^{2}}\left\vert i\right\rangle \left\langle j\right\vert ~,  \label{Action_Ph_Damp}
\end{equation}
i.e., we have, as expected, partial suppression of only the off-diagonal matrix elements of the state in the computational $Z$-basis. This
latter equation shows why this second definition of the phase damping channel for qudits reproduces the usual phase decoherence effect.

\section{Input-Output Fidelities}

\label{sec:fid}

Once we have defined the two kinds of phase damping channel for a qudit, the Weyl channel of Eqs.~(\ref{Kraus_mapD}) and (\ref{krausDweyl}) and
the conventional phase damping channel of Eq.~(\ref{Action_Ph_Damp}), we can consider their action upon our minimal phase code
$\{\widetilde{\left\vert 0\right\rangle },\widetilde{\left\vert 2k+1\right\rangle }\}$. Here, we encode a logical qubit into a qudit (with
dimension $D=4k+2$) by means of these codewords, and we analyze the effects of phase damping with and without error recovery. Such effects are
quantified in terms of fidelity of the output logical state with respect to the input. The results are then compared to the case where a
bare qubit is sent through the channel, i.e., when neither encoding nor decoding is performed.

\begin{figure}[tbph]
\vspace{-0.0cm}
\par
\begin{center}
\includegraphics[width=0.46\textwidth]{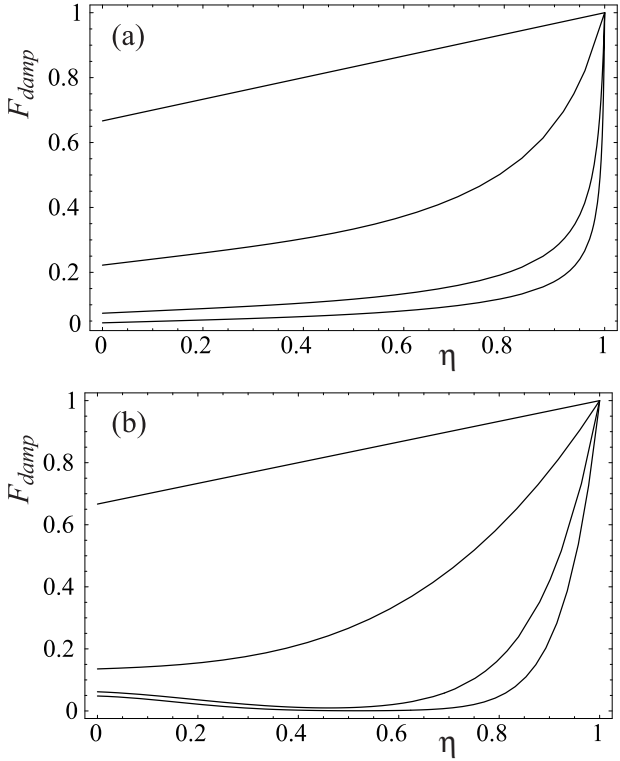}
\end{center}
\par
\vspace{-0.5cm}
\caption{Averaged input-output fidelities $%
F_{damp}$ in the case of a damped qudit. (a) refers to the conventional phase damping channel, while (b) refers to the Weyl channel. Fidelities
are plotted versus the damping parameter $\protect\eta $ and for different dimensions $D=30, 18, 6, 2$ from bottom to top.} \label{damped}
\end{figure}

Let us encode an arbitrary pure state $\cos (\theta /2)\left\vert
0\right\rangle +e^{i\phi }\sin (\theta /2)\left\vert 1\right\rangle $ of a
qubit into a coherent superposition of phase-codewords, i.e.,%
\begin{equation}
\left\vert \theta ,\phi \right\rangle =\cos \frac{\theta }{2}\widetilde{%
\left\vert 0\right\rangle }+e^{i\phi }\sin \frac{\theta }{2}\widetilde{%
\left\vert 2k+1\right\rangle }~.  \label{Logic_super}
\end{equation}%
In the computational basis, the logical state of Eq.~(\ref{Logic_super})
reads
\begin{equation}
\left\vert \theta ,\phi \right\rangle =\frac{1}{\sqrt{D}}\sum%
\limits_{l=0}^{D-1}\left[ \cos \frac{\theta }{2}+(-1)^{l}e^{i\phi }\sin
\frac{\theta }{2}\right] \left\vert l\right\rangle ~,
\end{equation}%
and the corresponding density operator is given by%
\begin{equation}
{\rho }(\theta ,\phi )=\left\vert \theta ,\phi \right\rangle \left\langle
\theta ,\phi \right\vert =\frac{1}{D}\sum\limits_{l,m=0}^{D-1}\Omega
_{lm}\left\vert l\right\rangle \left\langle m\right\vert ~,
\end{equation}%
where
\begin{eqnarray}
\Omega _{lm}:=[\cos \tfrac{\theta }{2}+(-1)^{l}e^{i\phi }\sin \tfrac{%
\theta }{2}][\cos \tfrac{\theta }{2}+(-1)^{m}e^{-i\phi }\sin \tfrac{\theta }{%
2}]~.\notag\\
\label{Omega}
\end{eqnarray}%
The effects of the phase damping Weyl channel and of the conventional phase damping channel on this logical state can be described with a
unified formalism. In fact, one can write the output state of the channel for the two cases as
\begin{equation}
\mathcal{E}\left[ {\rho }(\theta ,\phi )\right] =\frac{1}{D}%
\sum_{l,m=0}^{D-1}\Omega _{lm}f_r(\eta,l-m)\left\vert l\right\rangle \left\langle m\right\vert ~,  \label{output}
\end{equation}%
where $r=1$ refers to the conventional phase damping channel, $r=2$ to the Weyl channel and
\begin{eqnarray}
f_1(\eta,l-m) &=&\eta^{(l-m)^2} \\
f_2(\eta,l-m) &=&\left[\left(\frac{1-\eta}{2}\right)\omega^{(l-m)}+\left(\frac{1+\eta}{2}\right)\right]^{D-1}.
\end{eqnarray}
All the results can be expressed in terms of these two functions associated to each channel. In order to estimate the decoherence effects we
compute the fidelity between the input and output states
\begin{gather}
F_{damp}(\theta ,\phi ):=\left\langle \theta ,\phi \right\vert \mathcal{E}%
\left[ {\rho }(\theta ,\phi )\right] \left\vert \theta ,\phi \right\rangle
\notag \\
=\frac{1}{D^{2}}\sum_{l,m=0}^{D-1}\left\vert \Omega _{lm}\right\vert ^{2}f_r(\eta,l-m)~,
\end{gather}%
and, then, average this quantity over all the possible input states%
\begin{gather}
F_{damp}:=\frac{1}{4\pi }\int_{0}^{\pi }\sin \theta d\theta \int_{0}^{2\pi
}d\phi ~F_{damp}(\theta ,\phi )  \notag \\
=\frac{1}{3D^{2}}\sum_{l,m=0}^{D-1}\left[ 3+(-1)^{l-m}\right] f_r(\eta,l-m)~.  \label{F_Damp}
\end{gather}%
The behavior of the averaged fidelity is shown for the two cases in Fig.~\ref{damped}, where (a) refers to the $r=1$ conventional phase damping
channel and (b) to the $r=2$ Weyl channel. In both cases we note that the decoherence effect of the channel increases with the dimension $D$.

Let us now apply the recovery map of Eq.~(\ref{Rec_Channel_PhaseOLD}) to the
corrupted state of Eq.~(\ref{output}). The output (recovered) state is then
given by
\begin{equation}
\rho _{rec}(\theta ,\phi ):=\mathcal{\tilde{E}}_{R}\left\{ \mathcal{E}[{\rho
}(\theta ,\phi )]\right\} ~.
\end{equation}%
By inserting the matrix elements $\rho _{lm}=D^{-1}\Omega _{lm}f_r(\eta,l-m)$ of the corrupted state $\mathcal{E}[{\rho }(\theta ,\phi )]$ into
Eq.~(\ref{FI_Tilde_CompBase}) we get the corresponding $\tilde{\Phi}$
coefficients%
\begin{equation}
\tilde{\Phi}(x,y)=\frac{1}{D^{2}}\sum\limits_{l,m=0}^{D-1}\Omega _{lm}f_r(\eta,l-m)(-1)^{xl-ym}\Delta (l-m,D)~,  \label{FI_coeff}
\end{equation}%
which, in turn, must be substituted into Eq.~(\ref{Rec_Channel_PhaseOLD}) in
order to give the final explicit expression of the recovered state $\rho
_{rec}(\theta ,\phi )$. The input-output fidelity then becomes
\begin{gather}
F_{rec}(\theta ,\phi ):=\left\langle \theta ,\phi \right\vert \rho
_{rec}(\theta ,\phi )\left\vert \theta ,\phi \right\rangle  \notag \\
=\frac{1}{D^{2}}\sum\limits_{l,m=0}^{D-1}\Omega _{lm}f_r(\eta,l-m)\Delta
(l-m,D)  \notag \\
\times \left[ \cos ^{2}\frac{\theta }{2}+(-1)^{-m}\frac{\sin \theta }{2}%
e^{i\phi }\right.  \notag \\
\left. +(-1)^{l}\frac{\sin \theta }{2}e^{-i\phi }+(-1)^{l-m}\sin ^{2}\frac{%
\theta }{2}\right] ~,  \label{Fid_teta_fi}
\end{gather}%
and its average over all possible input states takes the form%
\begin{eqnarray}
&&F_{rec} :=\frac{1}{4\pi }\int_{0}^{\pi }\sin \theta d\theta \int_{0}^{2\pi
}d\phi ~F_{rec}(\theta ,\phi )  \notag \\
&&=\frac{1}{3D^{2}}\sum\limits_{l,m=0}^{D-1}\left[ 3+(-1)^{l-m}\right] f_r(\eta,l-m)\Delta (l-m,D)~,  \notag \\
&&  \label{Final_Fid}
\end{eqnarray}%
Note that the recovery fidelity of Eq.~(\ref{Final_Fid}) has the same form as the damped fidelity of Eq.~(\ref{F_Damp}) except for the presence
of the kernel-like term $\Delta (l-m,D)$ in Eq.~(\ref{Kernel}). This term formally takes the recovery operation into account and is therefore
responsible for the very different behavior of $F_{rec}$ and $F_{damp}$. Such a term is equal to $1$ only in the trivial case of a qubit ($%
D=2\Leftrightarrow k=0$) for which we have $F_{rec}=F_{damp}$, as is intuitively expected. In fact, in this case, the logical qubit is simply
encoded into another qubit and therefore remains unencoded.


Our qudit phase-code can be compared with
a $n$ qubit repetition code. Actually, for a given $D=4k+2$, the integer $n$ should be chosen as $odd\lceil \log{D} \rceil$, that is as the odd integer closest to $\log D$ from above.
For a $n$ qubit repetition code, starting from Eq.\eqref{krausNC}, we straightforwardly get \begin{eqnarray}
F_{rec}&=&\sum_{k=0}^{(n-1)/2}\left(
\begin{array}{c}
n\\
k
\end{array}\right)
\left(\frac{1+\eta}{2}\right)^{n-k}
\left(\frac{1-\eta}{2}\right)^{k}
\nonumber\\
&&+\frac{1}{3}\sum_{k=(n+1)/2}^{n}\left(
\begin{array}{c}
n\\
k
\end{array}\right)
\left(\frac{1+\eta}{2}\right)^{n-k}
\left(\frac{1-\eta}{2}\right)^{k}.\nonumber\\
\end{eqnarray}
Notice, that this result holds for both channels, and moreover for $n=2$ it corresponds to one qubit code, i.e. to unencoded qubit.

\begin{figure}[tbph]
\vspace{-0.0cm}
\par
\begin{center}
\includegraphics[width=0.46\textwidth]{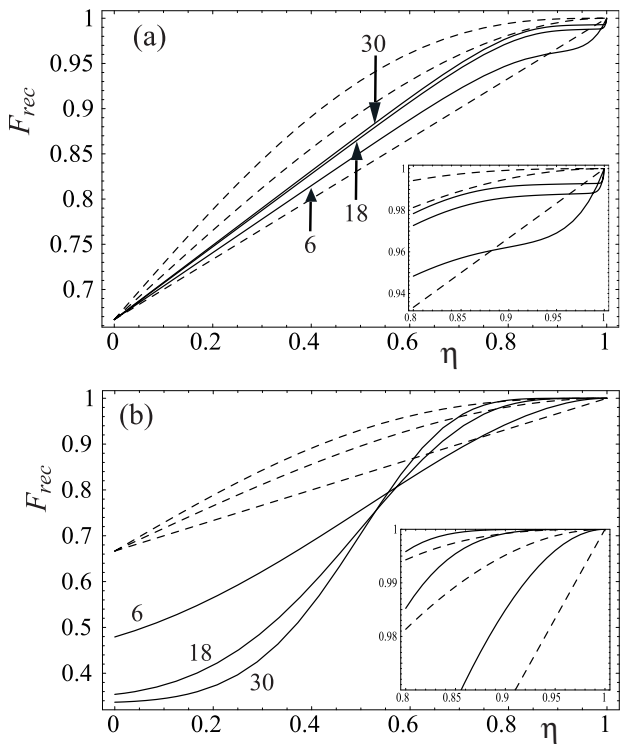}
\end{center}
\par
\vspace{-0.5cm}
\caption{Averaged input-output fidelities $%
F_{rec}$ in the case of a damped qudit in the presence of the recovery stage. (a) refers to the conventional phase damping channel, while (b)
refers to the Weyl channel. Fidelities are plotted versus the damping parameter $\protect\eta $ and for different dimensions $D=30, 18, 6$. Dashed lines refer, from top to bottom, to repetition code of dimension $32, 8, 2$ (using $5, 3, 1$ quibts respectively).
In (a) the qudit code is always worse than the repetition code, but it performs better than the unencoded case except for $\eta\to 1$ (see inset). In (b) the qudit code is effective at small phase damping $\eta \to 1$, where it can outperform the repetition code (see inset).}
\label{Recovered}
\end{figure}

The correcting power of our qudit phase-code is shown in Fig.~\ref{Recovered}, where the recovery fidelity of Eq.~(\ref{Final_Fid}) has been
plotted as a function of the channel decoherence parameter $\eta $ for several dimensions $D=4k+2$, for the two examples of phase-damped
channels, the conventional phase damping channel in Fig.~\ref{Recovered}a and the Weyl channel in Fig.~\ref{Recovered}b.

In Fig.~\ref{Recovered}a we see that in the case of conventional
phase damping the qudit code performs better and better for
increasing dimension and always outperforms the unencoded case
$D=2$ (lower dashed line in Fig.~\ref{Recovered}a). More
precisely, in both limits $\eta \rightarrow 1^{-}$ (weak damping)
and $\eta \rightarrow 0^{+}$ (strong damping) the correction
scheme does not depend upon $D$. However, the improvement with
respect to the unencoded transmission of the qudit is remarkable
in the intermediate regime (see Fig.~\ref{Recovered}a). We have
also made the comparison with the block-encoding. Actually, the
dotted lines from top to bottom in Fig.~\ref{Recovered}a refer to
repetition code of dimension $32, 8, 2$ (using $5, 3, 1$ qubits)
respectively. This code always outperforms the qudit code proposed
here for any $D$, showing that even though useful, our qudit codes
are not optimal in the case of the conventional phase damping
channel. This is however not surprising because the qudit code is
very different from repetition codes and it is not designed to
cancel errors at first order in the error probability as do the
latter codes.

The situation for the Weyl channel shown in Fig.~\ref{Recovered}b is more involved. In this case, the correcting power of the qudit code
increases for increasing $D$ \emph{only at small phase damping $\eta \to 1$}, while worsening for increasing dimension in the strong damping limit
$\eta \to 0$. This means that for a Weyl channel the qudit code outperforms the unencoded case $D=2$ only at large $\eta$ ($\eta > 0.7$) and
therefore it is useful only in the weak damping limit. In this limit however, contrary to what happens for the conventional phase damping
channel, the qudit code becomes particularly useful because it can outperform even the repetition code
(dotted lines from top to bottom in Fig.~\ref{Recovered}b refer to repetition code of dimension $32, 8, 2$  respectively).
In particular, while the qudit code of $D=6$ does not outperform the 3-qubit repetition code,
the qudit code of $D=18$ outperforms the 3-qubit repetition code,
and quite remarkably the
qudit code of $D=30$ outperforms the 5-qubit repetition code.

Finally, the qudit code turns into a non effective code while decreasing $\eta$ (worsening for increasing dimension),  because it is tailored to correct errors of weight up to $k$ while in such a limit
errors of higher weight becomes more and more probable.

\subsection{State-dependent fidelity}

Besides the average fidelity, it is also interesting to analyze the
state-dependent fidelity $F_{rec}(\theta ,\phi )$ of Eq.~(\ref{Fid_teta_fi}%
). Both phase damping channels act on the phase of the states and therefore in both cases the eigenstates of $Z$ are unaffected by decoherence,
as can be easily checked. It is interesting to see what happens to the encoded states in the presence of the recovery procedure of Sec. IID.
By construction, the two quantum codewords $\overline{\left\vert
+\right\rangle }:=\widetilde{\left\vert 0\right\rangle }$ and $\overline{%
\left\vert -\right\rangle }:=\widetilde{\left\vert 2k+1\right\rangle }$ of
our phase code are stabilized by $X^{2}$\ and connected by the logical phase
gate $\bar{Z}:=Z^{2k+1}$.
As a consequence they are significantly affected by both phase damping channels even in the presence of error correction. For the conventional
phase damping channel the two fidelities satisfy the simple relation
\begin{equation}
F_{rec}(0,0)=F_{rec}(\pi ,0)\rightarrow \tfrac{1}{2}~\text{for~}\eta
\rightarrow 0^{+}~,  \label{F_rec1}
\end{equation}%
i.e., they are completely dephased under the effect of strong phase damping. In the Weyl channel case the effect is even stronger and for strong
phase damping $\eta \to 0$ one has $F_{rec}(0,0)=F_{rec}(\pi ,0)\simeq  0$ at large enough $D$. However, one can still find an encoded basis
which is unaffected by phase damping in the presence of error correction. Such a basis is formed by the \emph{rotated} codewords given by the
two simple superpositions of the initial codewords,
\begin{eqnarray}
\left\vert \zeta _{0}\right\rangle &:&=\frac{\overline{\left\vert
+\right\rangle }+\overline{\left\vert -\right\rangle }}{\sqrt{2}}%
=(2k+1)^{-1/2}\sum\limits_{n=0}^{2k}\left\vert 2n\right\rangle ~, \\
\left\vert \zeta _{1}\right\rangle &:&=\frac{\overline{\left\vert
+\right\rangle }-\overline{\left\vert -\right\rangle }}{\sqrt{2}}%
=(2k+1)^{-1/2}\sum\limits_{n=0}^{2k}\left\vert 2n+1\right\rangle ~.
\end{eqnarray}%
One can easily check that these are the eigenstates of the logical phase
gate $\bar{Z}:=Z^{2k+1}$ and are connected by the single shift operator $X$,
i.e.,%
\begin{equation}
\bar{Z}\left\vert \zeta _{0}\right\rangle =\left\vert \zeta
_{0}\right\rangle ~,~\bar{Z}\left\vert \zeta _{1}\right\rangle =-\left\vert
\zeta _{1}\right\rangle ~,
\end{equation}%
and
\begin{equation}
X\left\vert \zeta _{0}\right\rangle =\left\vert \zeta _{1}\right\rangle
~,~X\left\vert \zeta _{1}\right\rangle =\left\vert \zeta _{0}\right\rangle ~.
\end{equation}%
In other words, these \emph{rotated} codewords $\{\left\vert \zeta _{0}\right\rangle, \left\vert \zeta_{1}\right\rangle\}$ behave like the $Z$-eigenstates $\left\vert
0\right\rangle $ and $\left\vert 1\right\rangle $ of a qubit. In fact, the corresponding fidelities satisfy, for both phase damping channels,
\begin{equation}
F_{rec}\left( \pm \pi /2,0\right) =1 ~, \quad\forall k,\eta ~,  \label{F_rec2}
\end{equation}%
which is exactly the behavior of the $Z$-basis eigenstates $\left\vert
0\right\rangle ,\left\vert 1\right\rangle $ under the action of phase
damping (i.e., they remain unchanged). However, let us remark that these
correspondences with the single qubit eigenstates hold if and only if the
qudit codewords are subject to error correction.

One can explicitly show that the two codewords $\{\left\vert \zeta _{0}\right\rangle, \left\vert \zeta _{1}\right\rangle\}$ are perfectly restored by error recovery for any
value of $\eta$ by making use of Eq.~(\ref{Rec_Channel_PhaseOLD}) and Eq.~(\ref{FI_Tilde_CompBase}), from which one can verify that
\begin{gather}
\tilde{\Phi}(0,0)=(-1)^{u}\tilde{\Phi}(0,2k+1)=(-1)^{u}\tilde{\Phi}(2k+1,0)
\notag \\
=\tilde{\Phi}(2k+1,2k+1)=1/2~,
\end{gather}%
which implies%
\begin{equation}
\mathcal{\tilde{E}}_{R}\left[ \mathcal{E}(\left\vert \zeta _{u}\right\rangle \left\langle \zeta _{u}\right\vert )\right] =\left\vert \zeta
_{u}\right\rangle \left\langle \zeta _{u}\right\vert ~.
\end{equation}%

As a final remark, note how the error correcting properties of the
orthogonal choices $\overline{\left\vert +\right\rangle },\overline{%
\left\vert -\right\rangle }$ and $\left\vert \zeta _{0}\right\rangle
,\left\vert \zeta _{1}\right\rangle $ are perfectly the same in the regime
of weak damping (where they correct up to $k$ phase-shifts with exactly the
same ability), while their performances dramatically split when the phase
damping becomes heavier and it is no-more reducible to the standard QEC
regime.

\section{Conclusion}

\label{sec:con}

In conclusion, we have addressed the problem of how to profitably exploit
the extra space available by embedding a quantum system into a
\textquotedblleft larger\textquotedblright\ one (qudit-encoding). Such an
approach can be useful from the point of view of the experimental
feasibility of quantum error correction schemes, since the dimension $D$ of
the encoding Hilbert space remains reasonably low. In particular, we have
considered the minimal $D$ which enables the construction of qudit-codes
able to restore a logical qubit in specific decoherence models. These
minimal codes are then proven to be efficient in protecting quantum
information against the detrimental effects of phase damping. This study
could shed further light into the role that Hilbert space dimensions play
in quantum error correction. The opposite problem of quantum data
compression could be considered in the same light. That is, data compression
from $\mathcal{H}_{2}^{\otimes n}$ to $\mathcal{H}_{D}$ with $2^{n}>D$, as
it has been considered in Refs.~\cite{GW03} and~\cite{Gr2}.

Possible experimental implementations of these codes and the corresponding recovery operations require the ability to efficiently implement the
generalized Pauli operators $X$ and $Z$ in an effective D-dimensional system. An interesting opportunity is provided by ring-shaped optical
lattices, which have been proposed as a possible quantum simulator of periodic one-dimensional quantum systems \cite{cata}. If we place a single
atom in a ring-shaped lattice with D sites, the ground states in each site are the basis states. As a consequence, the amplitude shift $X$ is
realized by tunneling, while the phase shift $Z$ could be realized by applying controlled local Stark shifts to the atom.

\acknowledgments

Authors thank Stojan Rebic and Markus Grassl for useful comments. S.P. also
thanks Gaetana Spedalieri for the strong encouragement, given while
finishing this work. This work was supported by the European Commission
through the Integrated Project FET/QIPC \textquotedblleft SCALA".


\begin{thebibliography}{99}
\bibitem{note1} Formally, a stabilizer code is perfect if all the
eigenvalues of the generators constitute valid syndromes for correcting an
error.
\bibitem{note2} Note that these operators are not Hermitian. In fact, they
derive from a \emph{particular }generalization of the Pauli operators, the
one which preserves the unitarity \cite{Knill}.
\end{thebibliography}
\end{document}